\documentclass{iau}
\usepackage{graphicx,natbib}
 
\def\Msun{{\mathrm M}_{\odot}}
\def\Msunyr{\mathrm {M}_{\odot}\,\mathrm {yr}^{-1}}

\title{Simulations on the survivability of Tidal Dwarf Galaxies}

\author[Ploeckinger, Recchi, Hensler \& Kroupa]{Sylvia Ploeckinger$^{1,2}$,Simone Recchi$^{1,3}$, Gerhard Hensler$^{1,4}$ \and Pavel Kroupa$^5$}

\affiliation{$^1$Department for Astrophysics, University of Vienna,\\ T\"urkenschanzstr. 17, 1180 Vienna, Austria \\
$^2$email: {\tt sylvia.ploeckinger@univie.ac.at} \\
$^3$email: {\tt simone.recchi@univie.ac.at} \\
$^4$email: {\tt gerhard.hensler@univie.ac.at} \\
$^5$Helmholtz-Institut fur Strahlen- und Kernphysik \\
Nussallee 14–16, D-53115 Bonn, Germany \\ email: {\tt pavel@astro.uni-bonn.de}}

\pubyear{2014}
\volume{309}
\jname{Galaxies in 3D acrsoss the Universe}
\editors{B. L. Ziegler, F. Combes, H. Dannerbauer, M. Verdugo, eds.}

\begin{document}

\maketitle

\begin{abstract}

We present detailed numerical simulations of the evolution of Tidal Dwarf Galaxies (TDGs) after they kinematically decouple from the rest of the tidal arm to investigate their survivability. Both the short-term (500 Myr) response of TDGs to the stellar feedback of different underlying stellar populations as well as the long-term evolution that is dominated by a time dependent tidal field is examined. All simulated TDGs survive until the end of the simulation time of up to 3 Gyr, despite their lack of a stabilising dark matter component.

\end{abstract}

\firstsection
\section{Introduction}

The formation of TDGs, as actively star-forming gaseous over-densities embedded in the tidal arms of interacting galaxies, is studied with observations in various wavelengths and
their formation is modelled within large-scale simulations of galaxy interactions \citep[see][for a review]{Duc2013}. Whereas they are easily identifiable when they are still connected to the tidal arm, their typical lifetimes remain under discussion. The contribution of ancient TDGs to the population of low-mass galaxies depends on both their formation rate as well as on their survivability. The estimates of dwarf galaxies (DGs) with a tidal origin cover a wide range from 6\% \citep{Kaviraj2012} to 100\% \citep{Okazaki2000}. Long-term chemo-dynamical simulations of TDGs do not only allow for a better estimate of their typical lifetimes but can also constrain the physical and
chemical properties of ancient TDGs.

\section{Results}
{\bf Early response to stellar feedback:} The initial conditions of the simulated TDGs represent a spherical, pressure supported gas cloud with cold over-densities that serve as the seeds for immediate star formation (SF). Within 20 Myr after the simulations start, the total SF rate of the TDGs reaches $5 \times 10^{-2} \, \Msunyr$. We study the response of dark matter (DM)-free DGs to different stellar feedback scenarios. In the low feedback case, the initial mass function (IMF) of each star cluster is truncated at a maximal star mass up to which at least a single star can be formed and that is dependent on the mass of the star cluster \citep[$m_{\mathrm{max}}-M_{\mathrm{ecl}}$ relation and IGIMF theory, for details see][]{Kroupa2003, Weidner2013}. For the high feedback case the IMF is assumed to be completely filled up to a maximal star mass of $120\,\Msun$ for every star cluster. In the first 300 Myr, the stellar feedback regulates the SF significantly stronger in the high feedback case, resulting in a 6 times lower stellar mass after 500 Myr than in the low feedback case ($8.35 \times 10^5 \,\Msun$ and $5.73 \times 10^6\,\Msun$, respectively). After 300 Myr, both simulated TDGs regulate their SF to a constant rate below $10^{-4}\,\Msunyr$. In both cases, the initially high SF does not lead to a disruption of the TDG \citep{Ploeckinger2014a}.

{\bf Long term evolution:} Starting from a smooth, warm, rotating gas cloud in the tidal field of a massive galaxy, we simulate the long-term evolution of TDGs. In order to carve out the effect of the tidal field, a comparison simulation of an identical but isolated TDG is performed. In the first Gyr the SF is comparable. After 1 Gyr, as the TDG in the tidal field is approaching the peri-center of its orbit, the TDG is compressed and subsequently the SF increases by more than 2 orders of magnitude, but both TDGs survive until the end of the simulation time at t = 3 Gyr \citep[see Fig. 1 and][subm.]{Ploeckinger2014b}.

\section{Conclusions}
We simulate pressure supported and rotating TDGs, with low and high stellar 
feedback, and exposed them to ram pressure and a tidal field. All simulated TDGs ($Z = 0.1 - 0.3 \,\mathrm{Z}_{\odot}$) survive until the end of the simulation (up to 3 Gyr). This serves as an additional sign that TDGs can turn into long-lived objects,
complementary to indications from observations \citep{Dabringhausen2013, Duc2014} and simulations of galaxy interactions \citep[e.~g.][]{Bournaud2006,Hammer2010,Fouquet2012}.

\begin{figure}
\centering
\includegraphics[width=0.82\columnwidth]{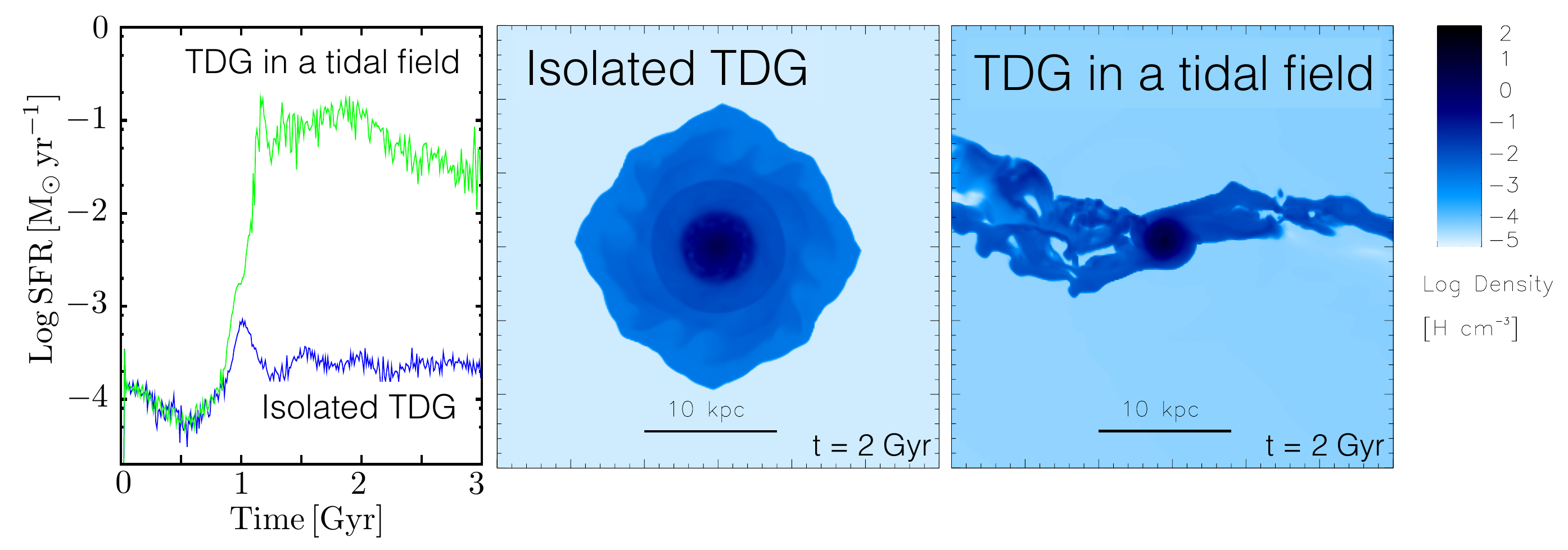} 
\caption{Comparison of the SFRs of initially identical TDGs with and without a tidal field (left panel). A face-on density slice is shown for the isolated TDG (middle panel) as well as the TDG in a tidal field (right spanel) at the peri-center passage of the orbiting TDG (t = 2 Gyr).}\label{fig:fig1}
\end{figure}

\bibliographystyle{mn2e}
\bibliography{ploeckinger_galaxy3d}

\end{document}